\documentclass[a4paper,11pt]{article}
\usepackage{pos}
\usepackage{bm}

\title{Single static-quark system above $T_c$ investigated by energy-momentum tensor in SU(3) Yang-Mills theory
\\ \vspace*{-69mm}\hspace{12.3cm} \small{\texttt{J-PARC-TH-0257}} \vspace*{67mm}}
\ShortTitle{Single static-quark system above $T_c$}

\author*[a,b]{Masakiyo Kitazawa}
\author[a]{Ryosuke Yanagihara}
\author[a]{Masayuki Asakawa}
\author[c]{Tetsuo Hatsuda}

\affiliation[a]{Department of Physics, Osaka University, Toyonaka, Osaka 560-0043, Japan}

\affiliation[b]{J-PARC Branch, KEK Theory Center, Institute of Particle and Nuclear Studies, KEK, 203-1, Shirakata, Tokai, Ibaraki, 319-1106, Japan}

\affiliation[c]{RIKEN Interdisciplinary Theoretical and Mathematical Sciences Program (iTHEMS), RIKEN, Wako 351-0198, Japan}

\emailAdd{kitazawa@phys.sci.osaka-u.ac.jp}

\abstract{
  We investigate the distribution of energy-momentum tensor (EMT) around a static quark in the deconfined phase of SU(3) Yang-Mills theory. The EMT defined through the gradient-flow formalism is used for the numerical analysis of the EMT distribution around the Polyakov loop with the continuum extrapolation. Using EMT, one can study the mechanical distortion of the color gauge field induced by the static charge. We find substantial separation in the absolute values of the EMT eigenvalues which is not observed in Maxwell theory. The separation grows as temperature is lowered toward the critical temperature. The lattice data also indicate the thermal screening at long distance and the perturbative behavior at short distance.
}

\FullConference{%
 The 38th International Symposium on Lattice Field Theory, LATTICE2021
  26th-30th July, 2021
  Zoom/Gather@Massachusetts Institute of Technology
}

\begin{document}
\maketitle

\section{Introduction}

One of the fundamental methods to explore a system described by
the quantum field theory is to introduce test charge(s) and analyze
the response of the system.
In this proceedings, we investigate the response of SU(3) Yang-Mills (YM)
theory against the existence of a static quark in the deconfined phase
in lattice numerical simulations~\cite{Yanagihara:2020tvs}.
To study the local distortion of the color field induced by
the static charge we use the energy-momentum tensor (EMT),
$T_{\mu\nu}(x)$.

Recently, successful analysis of the correlation functions of EMT
operators in lattice numerical simulations has become possible
thanks to the small-flow time expansion (SF$t$X) method~\cite{Suzuki:2013gza,Makino:2014taa}
based on the gradient flow~\cite{Narayanan:2006rf,Luscher:2010iy,Luscher:2011bx}.
Its application to the analysis of thermodynamics in SU(3) YM
theory~\cite{Asakawa:2013laa,Kitazawa:2016dsl,Iritani:2018idk}
and QCD with fermions~\cite{Taniguchi:2016ofw,Taniguchi:2020mgg}
shows that this method defines the EMT operator properly and 
is effective in suppressing statistical errors in numerical simulations.
The EMT operator defined in this method has been applied for the analysis
of various systems~\cite{Kitazawa:2017qab,Yanagihara:2018qqg,Kitazawa:2019otp,Yanagihara:2019foh}.
We use this method for the measurement of $T_{\mu\nu}(x)$ in this study.

\section{Energy-momentum tensor and stress tensor}

The EMT $T_{\mu\nu}(x)$ is related to the
stress tensor $\sigma_{ij}$ as
\begin{align}
 \sigma_{ij} = - T_{ij} \qquad (i,j = 1,2,3).
\end{align}
The force per unit area ${\cal F}_i$
is given by 
${\cal F}_i =  \sigma_{ij}n_j =  - T_{ij} n_j$
with the normal vector $n_j$.
The principal axes $n_j^{(k)}$ and the corresponding eigenvalues
$\lambda_k$
of the stress tensor are obtained by the eigenequation
\begin{align}
 T_{ij}n_j^{(k)}=\lambda_k n_i^{(k)} \quad (k=1,2,3),
 \label{eq:EV}
\end{align}
where $\lambda_k <0 $ ($\lambda_k >0 $) means that 
neighboring volume elements pull (push) each other on the surface
with $n_i^{(k)}$.

A system with a single static source has spherical symmetry.
We thus employ the spherical coordinate system $(r,\theta, \varphi)$
in the following.
The EMT is diagonalized in this coordinate system as
\begin{align}
  T_{\gamma\gamma'}(\bm{x})&=
  \mathrm{diag}(T_{44}(r), T_{rr}(r), T_{{\theta\theta}}(r))
 \label{eq:stress_separate}
\end{align}
where $\gamma,\gamma'={4,r,\theta}$.
Due to the spherical symmetry, 
the azimuthal and polar components degenerate,
$T_{\varphi\varphi}(r)= T_{\theta\theta}(r)$,
so that only independent components are shown
in Eq.~(\ref{eq:stress_separate}).

In Maxwell theory,
EMT is given by the Maxwell stress-energy tensor
$ T^\mathrm{Maxwell}_{\mu\nu}
=  {F}_{\mu\rho } {F}_{\nu \rho} - \frac{1}{4} \delta_{\mu\nu}
{F}_{\rho\sigma} {F}_{\rho\sigma}$
with the field strength ${F}_{\mu\nu}$.
The EMT in a static system is given by
\begin{align}
T_{\gamma\gamma'}^{\rm Maxwell}
=\frac{1}{2}\mathrm{diag}(-\vec{E}^2, -\vec{E}^2, \vec{E}^2),
\label{eq:Maxwell}
\end{align}
with  ${E}_i(\bm{x})$ being  the electric field.

\section{Gradient flow}

In this study we define the local EMT operator $T_{\mu\nu}(x)$
using the small flow-time expansion (SF$t$X) method~\cite{Suzuki:2013gza,Makino:2014taa}
based on the gradient flow~\cite{Narayanan:2006rf,Luscher:2010iy,Luscher:2011bx}.
We consider the pure SU(3) YM gauge theory in Euclidean
space defined by the action,
\begin{align}
 S_\mathrm{YM} 
 = \frac{1}{4g_0^2}\int d^4x\,G_{\mu\nu}^a(x)G_{\mu\nu}^a(x),
 \label{eq:flow_action}
\end{align}
with the bare gauge coupling $g_0$ and the field strength $G_{\mu\nu}^a(x)$.
The gradient flow in this theory is defined through the evolution
equation of the gauge field called the
flow equation~\cite{Luscher:2010iy,Luscher:2011bx},
\begin{align}
 \frac{dA_\mu^a(t,x)}{dt} 
 = -g_0^2\frac{\delta S_\mathrm{YM}(t)}{\delta A_\mu^a(t,x)},
 \label{eq:GF}
\end{align}
with the gauge field $ A_{\mu}^a(x) $ and the initial condition 
$A_\mu^a(t=0,x)=A_\mu^a(x)$.

Using Eq.~(\ref{eq:GF}), 
the properly renormalized EMT operator is defined with the small $t$
expansion~\cite{Suzuki:2013gza}:
\begin{eqnarray}
 T_{\mu\nu} (x)
  &= &\lim_{t\to0} T_{\mu\nu}(t,x) , \label{eq:t} \\ 
 T_{\mu\nu}(t,x)  &= &
 c_1(t) U_{\mu\nu}(t,x)
 + 4c_2(t) \delta_{\mu\nu}
  \left[E(t,x)-\left\langle E(t,x)\right\rangle_0 \right] ,
 \label{eq:T}
\end{eqnarray}
where $\langle E(t,x) \rangle_0$   is the vacuum expectation value.
The dimension-four gauge-invariant operators 
on the right-hand side of Eq.~(\ref{eq:T}) are
given by \cite{Suzuki:2013gza}
\begin{align}
  E(t,x) &= \frac{1}{4} G_{\mu\nu}^a(t,x)G_{\mu\nu}^a(t,x),
  \label{eq:E} 
  \\
  U_{\mu\nu}(t,x) &= G_{\mu\rho}^a (t,x)G_{\nu\rho}^a (t,x)
  - \delta_{\mu\nu} E(t,x).
  \label{eq:U}
\end{align}
For the coefficients $c_1(t)$ and $c_2(t)$ in Eq.~(\ref{eq:T}), we use the 
two-loop perturbative coefficients~\cite{Harlander:2018zpi,Iritani:2018idk}.

\section{Lattice setup}

A static quark $Q$ on the lattice is represented by the Polyakov loop $\Omega$.
The expectation value of $T_{\mu\nu}(x)$ around a static quark $Q$
at the origin is given by
\begin{align}
 \langle T_{\mu\nu}(t,x) \rangle_Q
 = \frac{\langle  T_{\mu\nu}(t,x)\mathrm{Tr}\Omega(\bm{0})\rangle}
 {\langle \mathrm{Tr}\Omega(\bm{0}) \rangle}
 -\langle  T_{\mu\nu}(t,x)\rangle,
 \label{eq:cor_pol_emt0}
\end{align}
with the Polyakov loop $\Omega(\bm{0})$ located at the origin.
Note that Eq.~(\ref{eq:cor_pol_emt0}) is well-defined only when
the $Z_3$ symmetry in SU(3) YM theory is spontaneously broken.
When the $Z_3$ symmetry is not broken, both numerator and denominator
of the first term on the right-hand side in Eq.~(\ref{eq:cor_pol_emt0})
vanish from symmetry.
We thus focus on the system in the deconfined phase
above the critical temperature $T_c$ in the present study.

Numerical simulations are performed with the Wilson gauge action 
for four different temperatures $1.20T_c,\,1.44T_c,\,2.00T_c$, and $2.60T_c$
with the aspect ratio $N_s/N_\tau=4$.
For each temperature we perform numerical simulations for five lattice
spacings to perform the continuum extrapolation.
Among them, we finally excluded the coarsest lattices from
the analysis~\cite{Yanagihara:2020tvs}.
We employ the Wilson gauge action for the flow action $S_\mathrm{YM}(t)$
in Eq.~(\ref{eq:GF}) and the clover-type
representation for the field strength $G_{\mu\nu}(t,x)$.
In the measurement of the Polyakov loop,
we apply the multi-hit procedure to suppress the statistical noise~\cite{Yanagihara:2020tvs}.

The renormalized EMT distribution around $Q$ is obtained after 
taking the double extrapolation,
\begin{align}
 \langle T_{\mu\nu}(x) \rangle_Q
 = \lim_{t \rightarrow 0} \lim_{a \rightarrow 0}
 \langle T_{\mu\nu}(t,x) \rangle_Q.
 \label{eq:w-ext}
\end{align}
In our actual analysis, we extract the EMT distribution
by fitting the lattice data obtained at the lattice spacing $a$
with the following functional form~\cite{Kitazawa:2016dsl,Kitazawa:2017qab}:
\begin{align}
 \langle T_{\mu\nu}(t,x) \rangle_Q
 = \langle T_{\mu\nu}(x) \rangle_Q
 + b_{\mu\nu}{(t)}a^2 + c_{\mu\nu}t + d_{\mu\nu}t^2, 
 \label{eq:double_lim}
\end{align}
where the contributions from discretization effects ($b_{\mu\nu}$)  
as well as the  dimension-six and -eight 
operators ($c_{\mu\nu}$ and $d_{\mu\nu}$) 
are considered.

\begin{figure}[t]
 \centering
 \includegraphics[width=0.38\textwidth, clip]{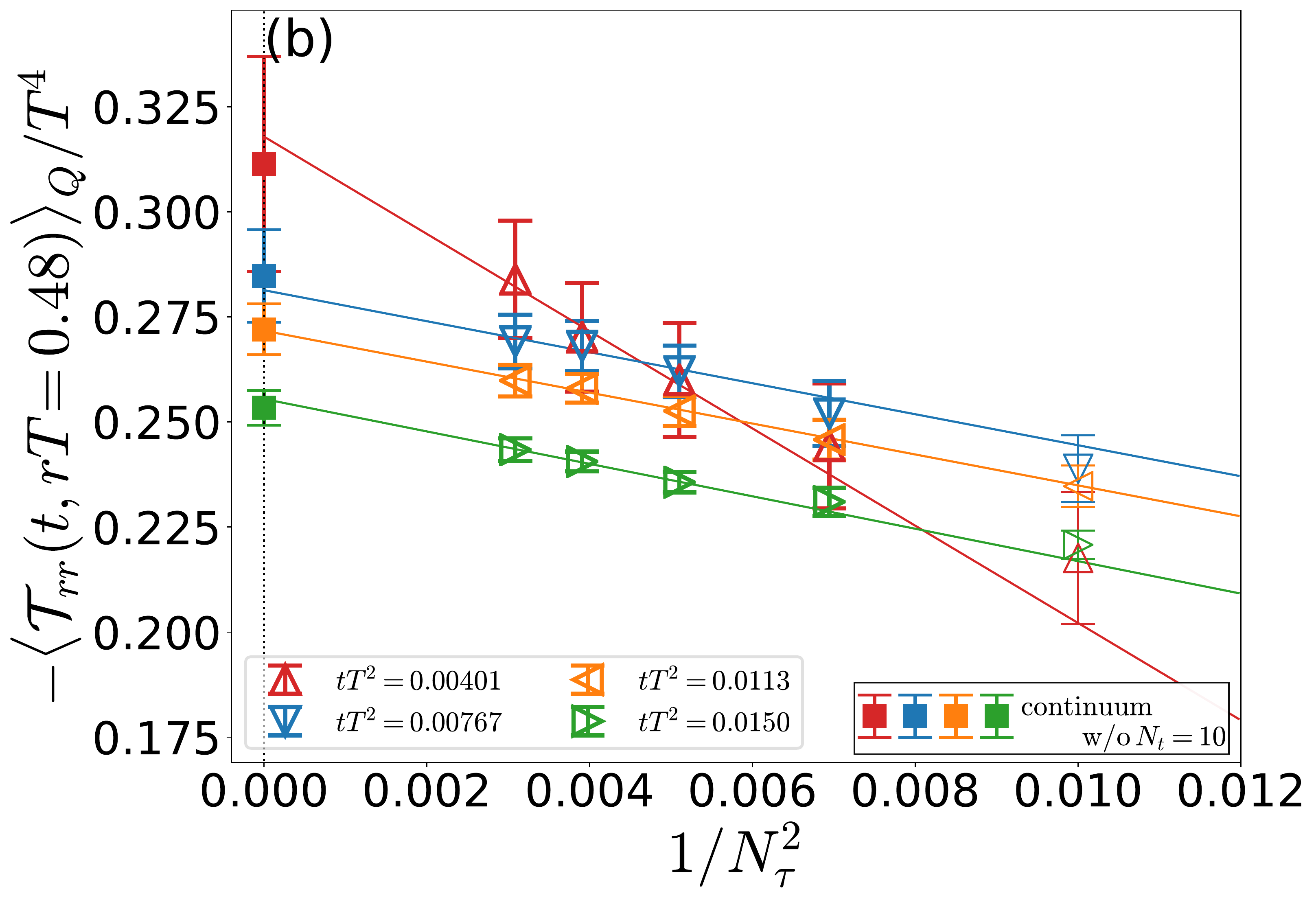}
 \includegraphics[width=0.3\textwidth, clip]{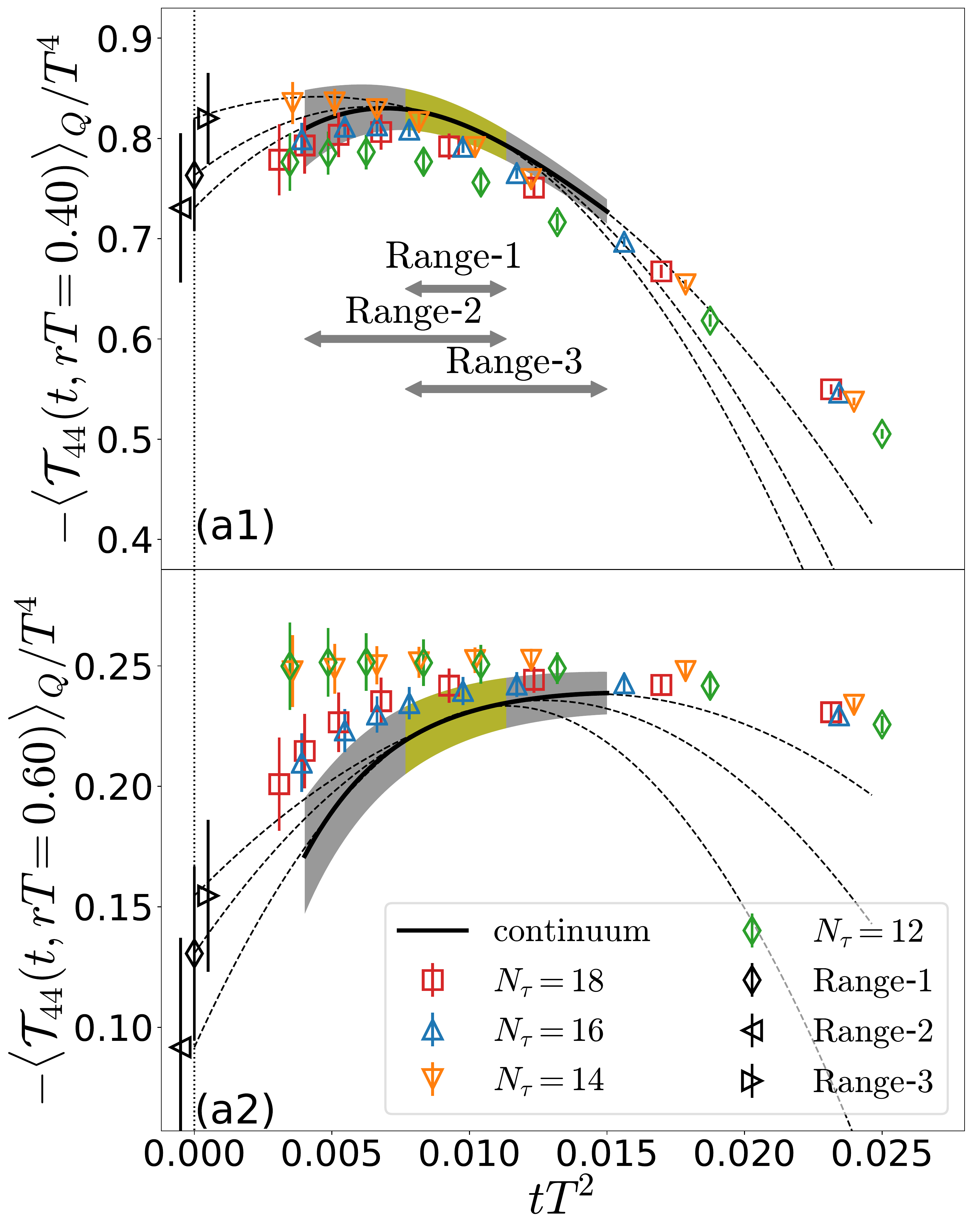}
 \includegraphics[width=0.3\textwidth, clip]{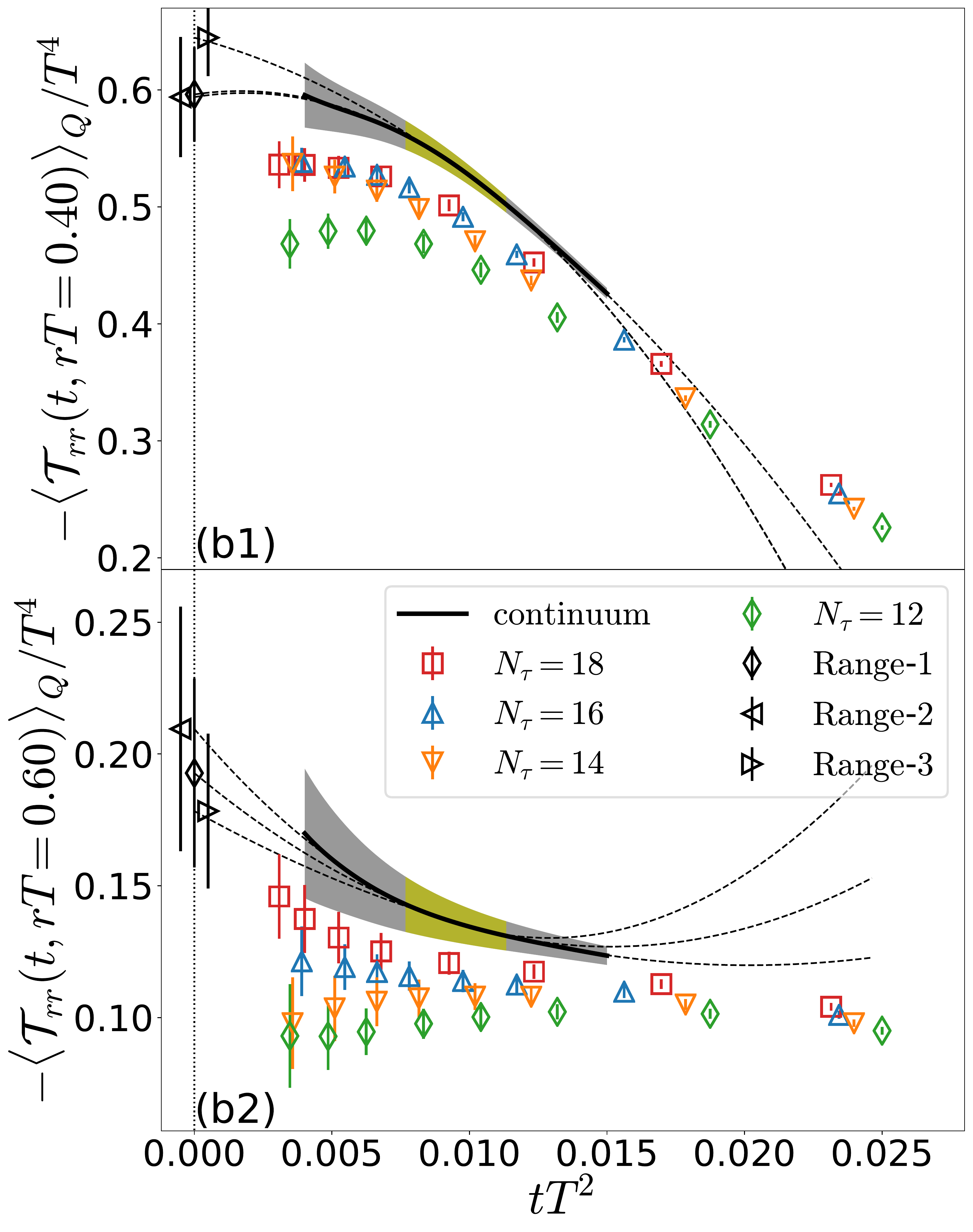}
 \caption{
   Left: $-\langle T_{rr}(t, rT) \rangle_Q/T^4$
   at $T/T_c=1.44$ and $rT=0.48$
   as functions of $1/N_\tau^2=a^2T^2$~\cite{Yanagihara:2020tvs}.
   The continuum extrapolation is shown by the solid lines and 
   the filled symbols.
   Middle and right:
   $-\langle T_{44}(t, rT) \rangle_Q/T^4$ and 
   $-\langle T_{rr}(t, rT) \rangle_Q/T^4$ as functions of $tT^2$
   at $rT=0.40$ (top) and $rT=0.60$ (bottom).
   Open symbols denote 
   $\langle T_{44}(t, rT) \rangle_Q/T^4$
   for each $a$.
   The black solid line is the continuum-extrapolated result.
   The dotted lines show the fitted results of the continuum result
   with Range-1, 2, and 3.
   The black symbols at $tT^2=0$ are the results of the $t\to0$ extrapolation.
 }
 \label{fig:t_to_0}
\end{figure}

To carry out the double extrapolation Eq.~(\ref{eq:double_lim}),
we first take the continuum limit with fixed $t$,
and then carry out the $t\to0$ extrapolation.
In the left panel of Fig.~\ref{fig:t_to_0}, we show 
$-\langle T_{rr}(t, rT) \rangle_Q$
at $rT=0.48$ as a function of $1/N_\tau^2=(aT)^2$ 
for four values of $tT^2$.
In the panel, the fitting results of the continuum extrapolation
according to Eq.~(\ref{eq:double_lim}) at fixed $t$ 
are  shown by the solid lines
together with the results of the continuum limit 
on the vertical dotted line at $1/N_\tau^2=0$.

In the middle and right panels of Fig.~\ref{fig:t_to_0},
we show the values of $\langle T_{44}(t, rT) \rangle_Q$ and
$\langle T_{rr}(t, rT) \rangle_Q$ 
at $rT=0.40$ and $0.60$ for each lattice spacing.
The results of the continuum limit 
are denoted by the black solid lines with the gray
statistical error band.
Using the continuum-extrapolated result, 
we take $t\rightarrow0$ extrapolation
with Eq.~(\ref{eq:double_lim}) at $a=0$.
For the fitting, we employ three fitting ranges of $t$, Range-1, 2, 3,
shown in
the middle panel, which are chosen within the range of $t$ that avoids
over-smearing due to the gradient flow and at the same time
the discretization effects are well suppressed.

\section{Numerical results}

\begin{figure}
 \centering
 \includegraphics[width=0.38\textwidth, clip]{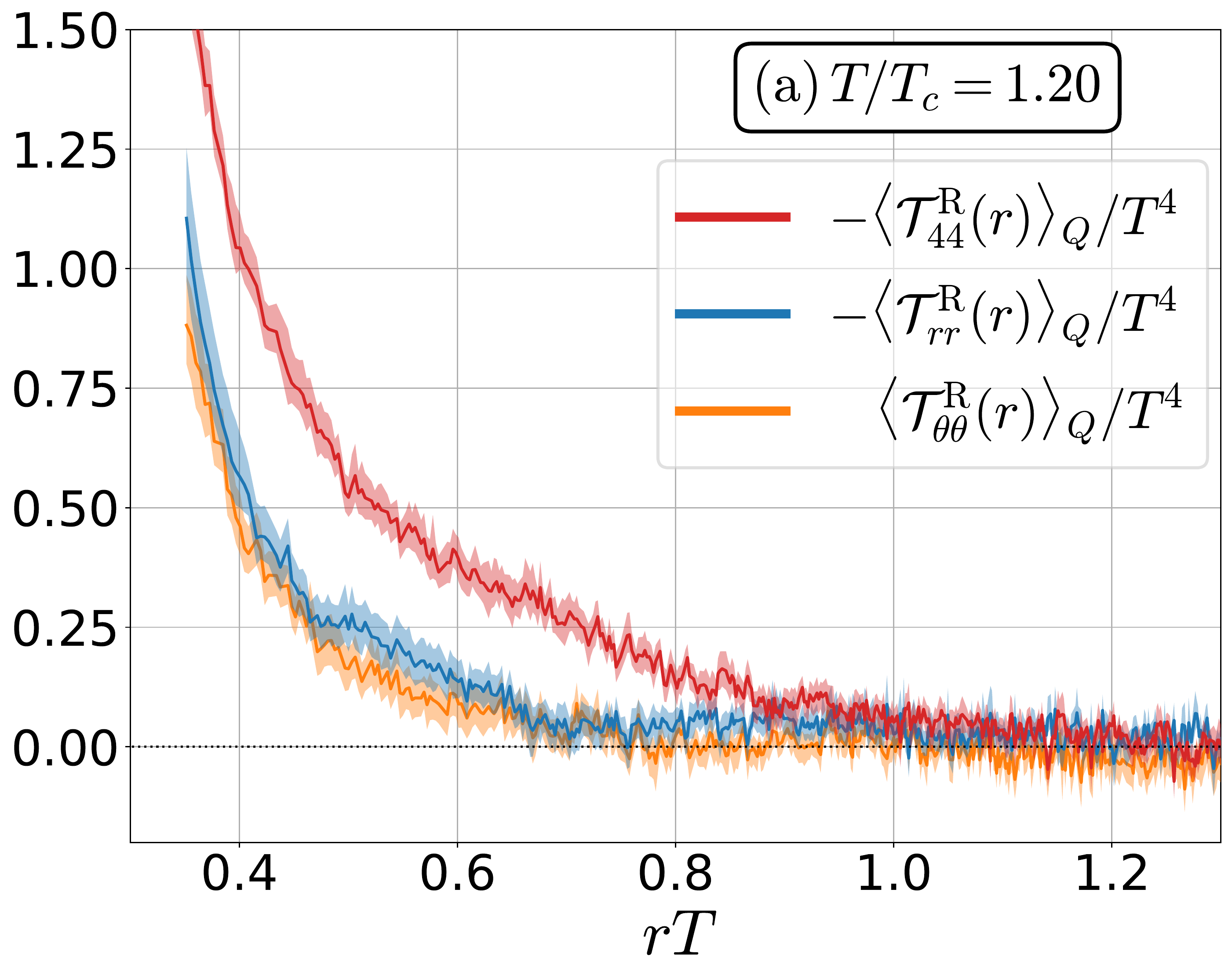}
 \includegraphics[width=0.38\textwidth, clip]{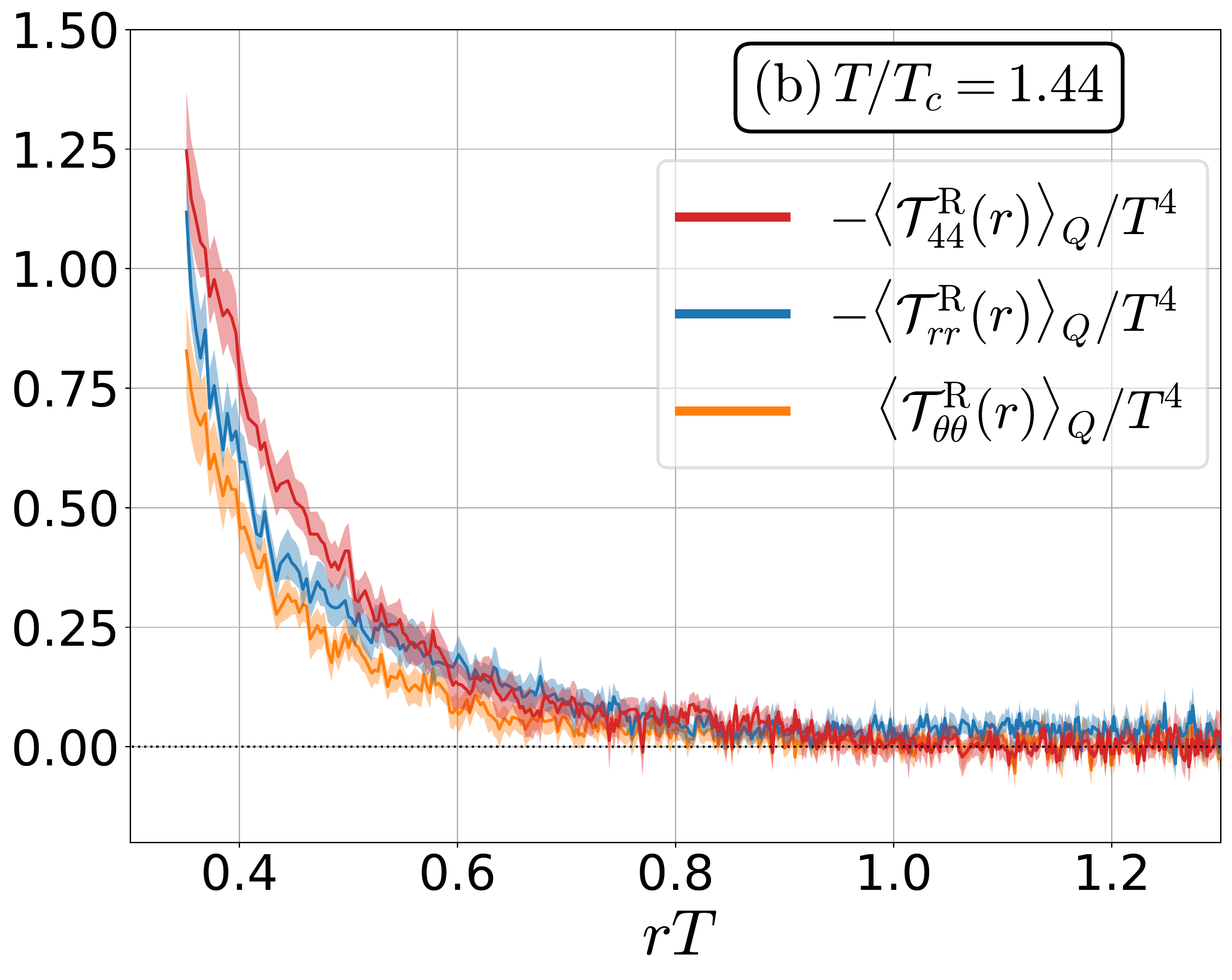}
 \\
 \includegraphics[width=0.38\textwidth, clip]{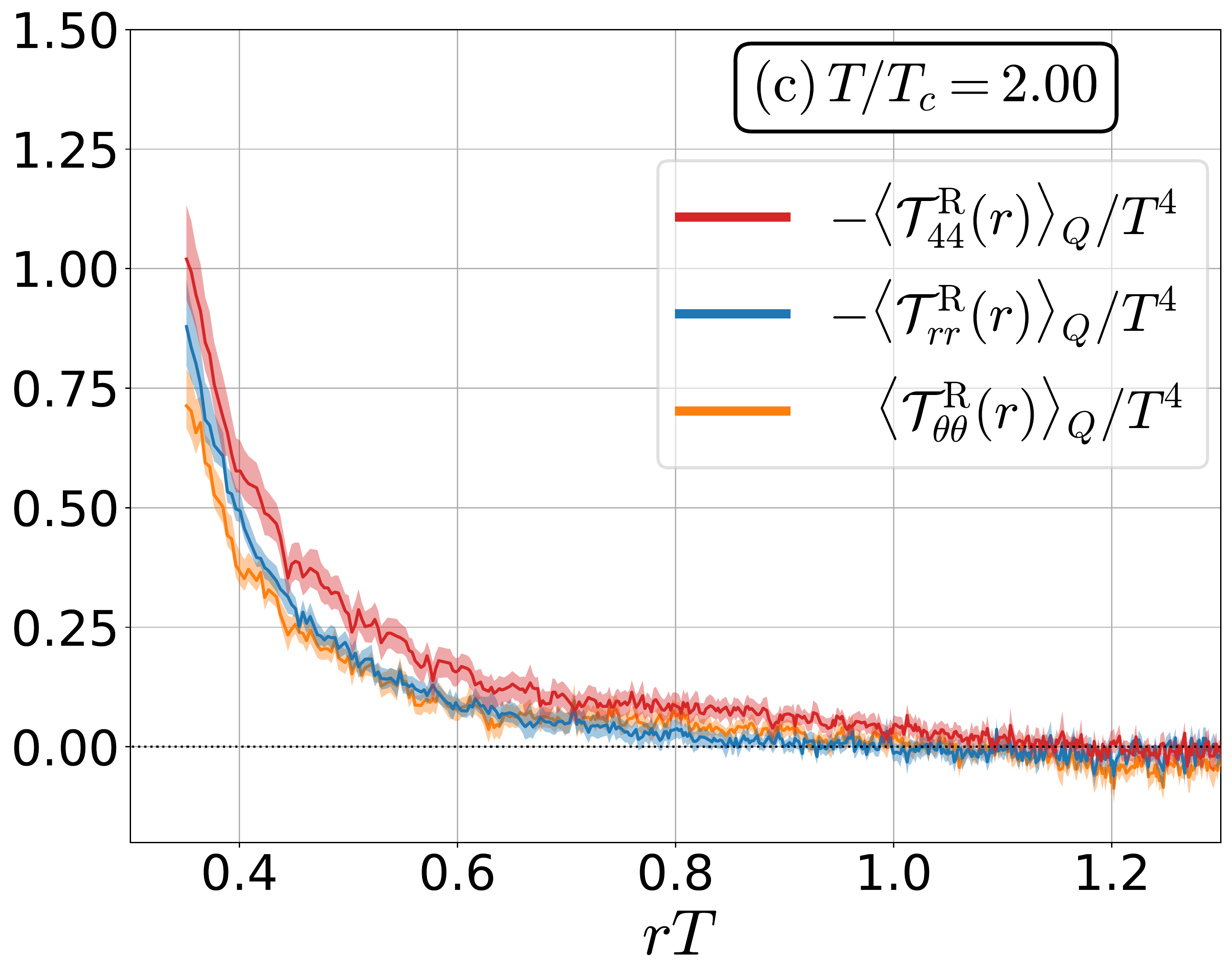}
 \includegraphics[width=0.38\textwidth, clip]{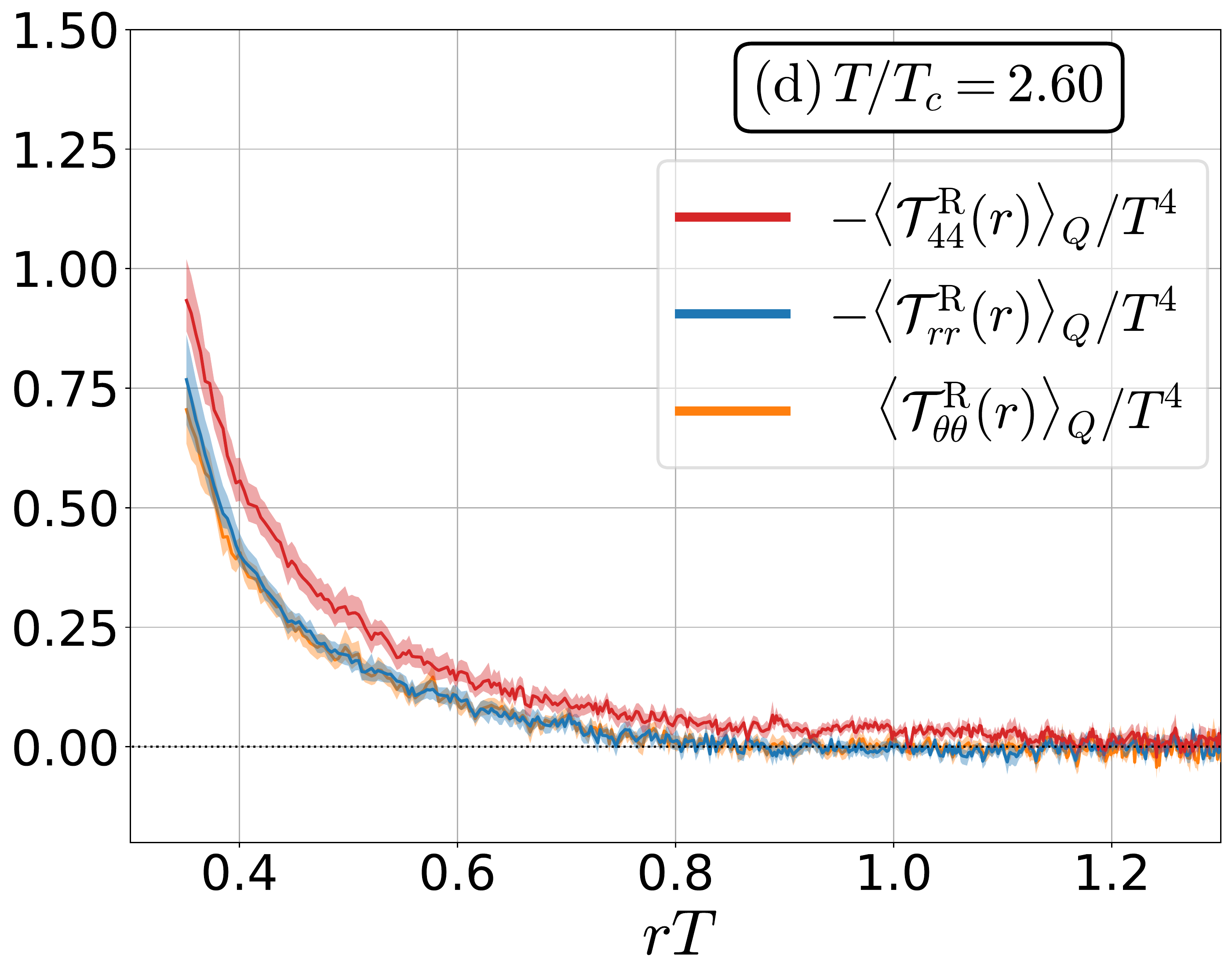}
 \caption{
 EMT distribution 
 $( -\langle T_{44} (r)\rangle_{Q},
 -\langle T_{rr} (r)\rangle_{Q},
 \langle T_{\theta \theta} (r)\rangle_{Q} )$
 as functions of $rT$
 after the double extrapolation~\cite{Yanagihara:2020tvs}.
 }
 \label{fig:each}
\end{figure}

In Fig.~\ref{fig:each}, we show the dimensionless EMT, 
$-\langle T_{44} (r)\rangle_{Q}/T^4$,
$-\langle T_{rr} (r)\rangle_{Q}/T^4$, and 
$\langle T_{\theta \theta} (r)\rangle_{Q} /T^4$,
as functions of $rT$. 
The error bands include both the statistical and systematic errors.
We find that  $-\langle T_{44} (r)\rangle_{Q}$, 
$-\langle T_{rr} (r)\rangle_{Q}$, and 
$\langle T_{\theta \theta} (r)\rangle_{Q}$ 
are all positive for $rT\lesssim1$ and
decrease rapidly with increasing $r$.
These signs are the same as those of 
the EMT in the Maxwell theory Eq.~(\ref{eq:Maxwell}).
The positive sign of $-\langle T_{44} (r)\rangle_{Q}$
means that the energy density is positive,
while the signs of $-\langle T_{rr} (r)\rangle_{Q}$ and 
$\langle T_{\theta \theta} (r)\rangle_{Q}$ show
that the volume element receives
pulling and pushing forces along the longitudinal and transverse
directions, respectively.
Figure~\ref{fig:each} indicates that 
the absolute values of the spatial components 
$|\langle T_{rr} (r)\rangle_{Q}|$ and 
$|\langle T_{\theta \theta} (r)\rangle_{Q}|$
are degenerated within the error for 
all temperatures.
On the other hand, 
$|\langle T_{44}  (r)\rangle_{Q}|$ is larger than 
the spatial components
especially at lower temperature.
This is in contrast to the degenerate magnitude of all components 
in the Maxwell stress Eq.~(\ref{eq:Maxwell}).
It is interesting to note that such a separation is 
also obtained in the thermal perturbation theory~\cite{Berwein}.

\begin{figure}[t]
 \centering
 \includegraphics[width=0.38\textwidth, clip]{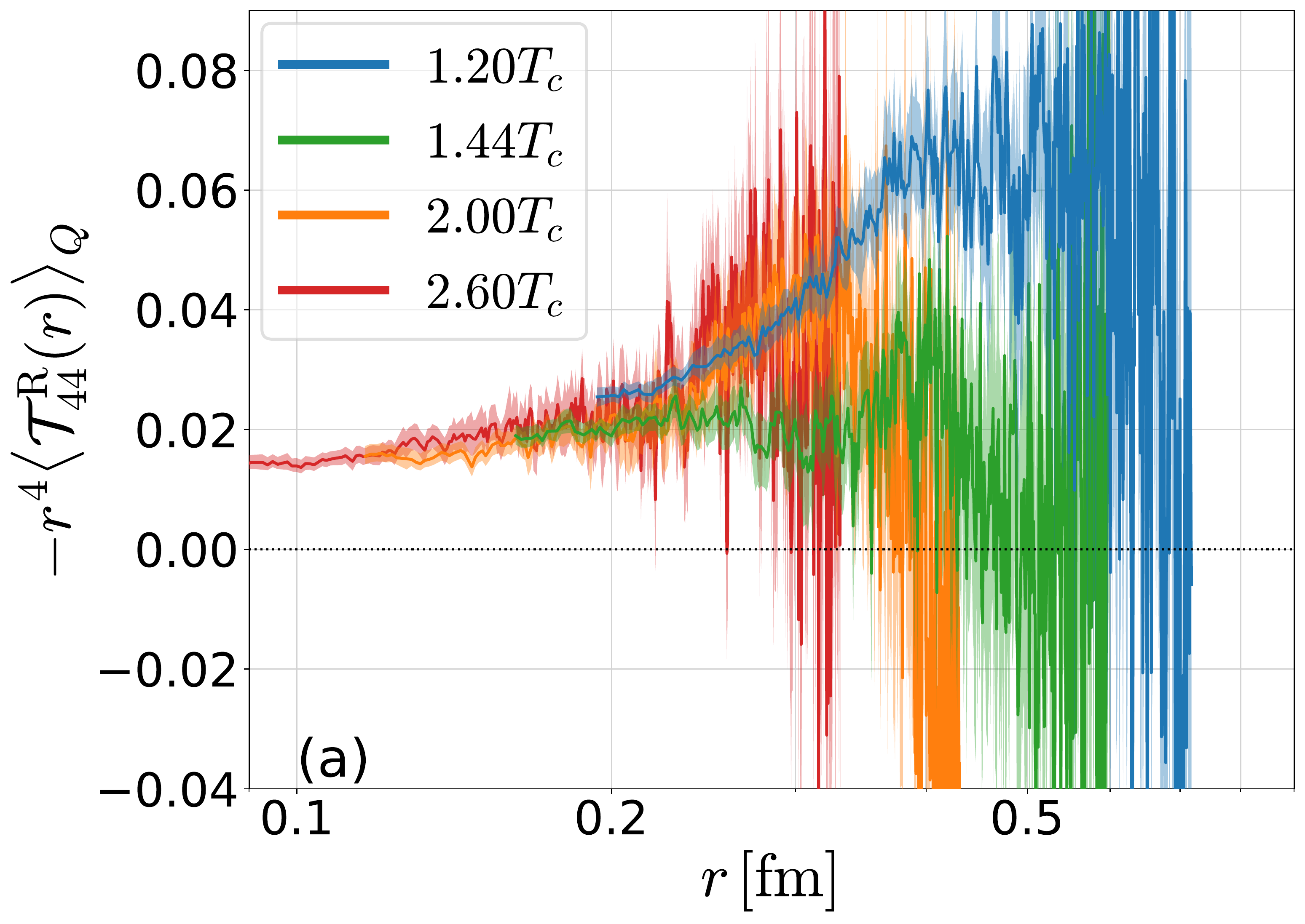}  
 \includegraphics[width=0.38\textwidth, clip]{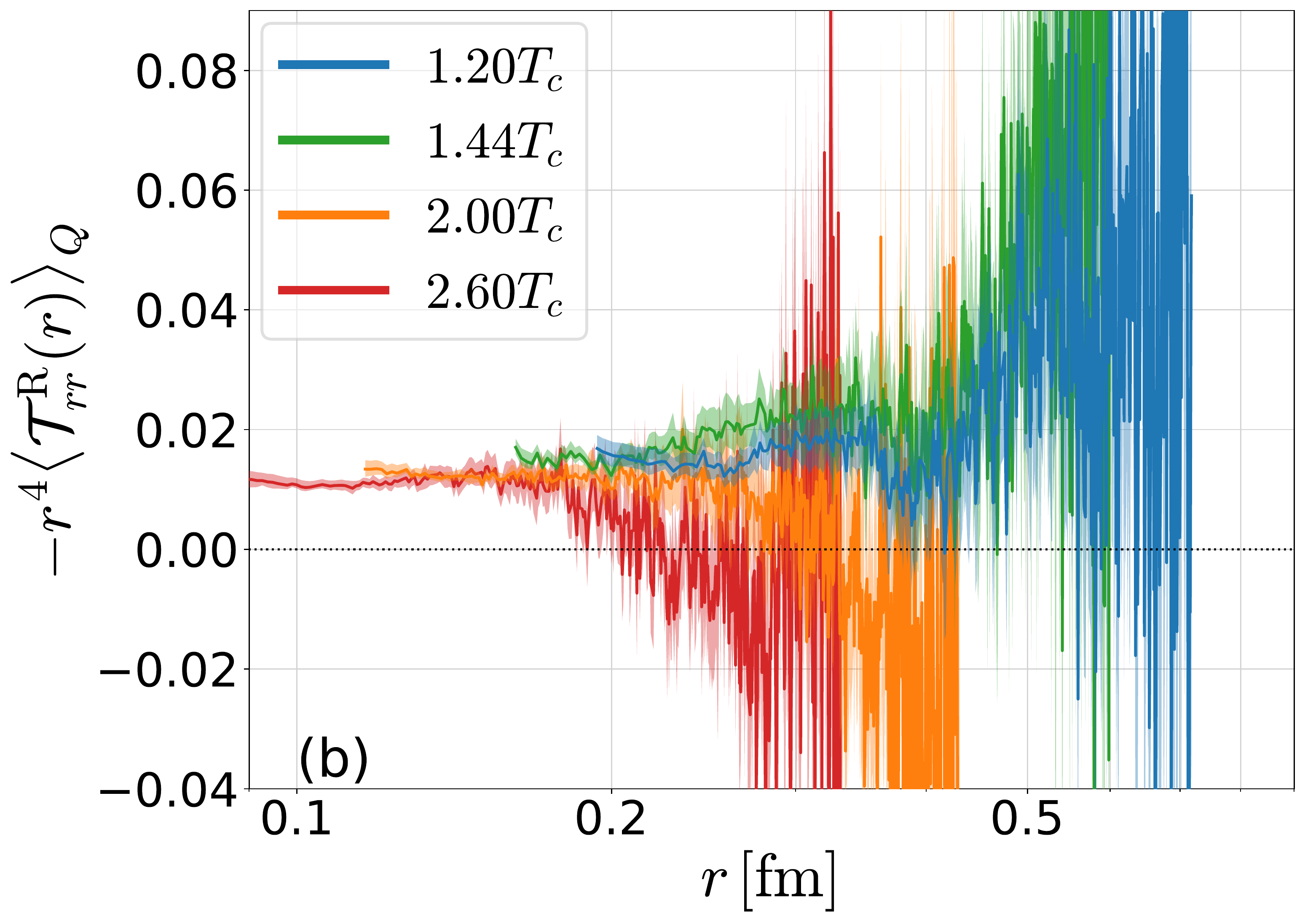}
 \\
 \includegraphics[width=0.38\textwidth, clip]{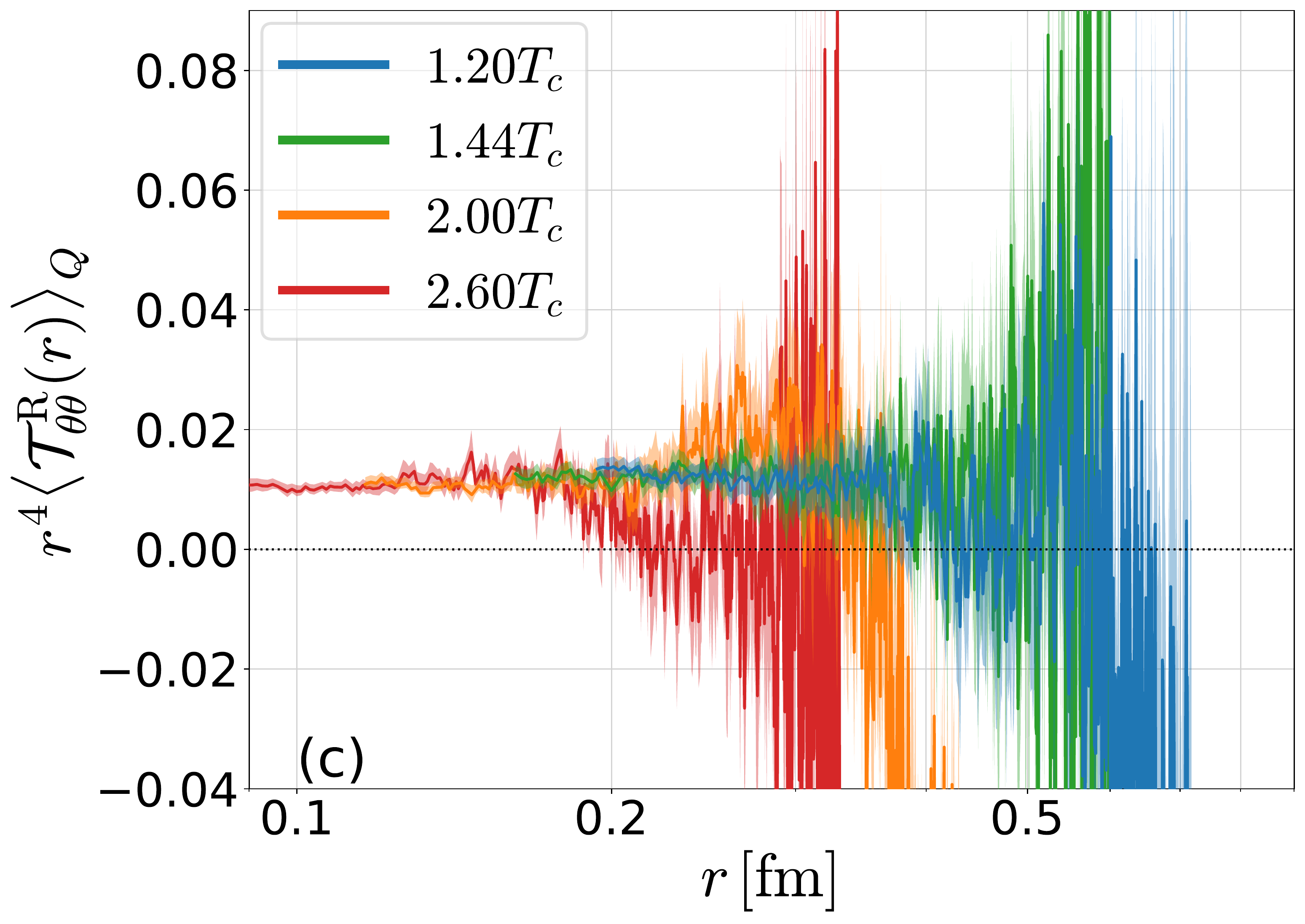}
 \includegraphics[width=0.38\textwidth, clip]{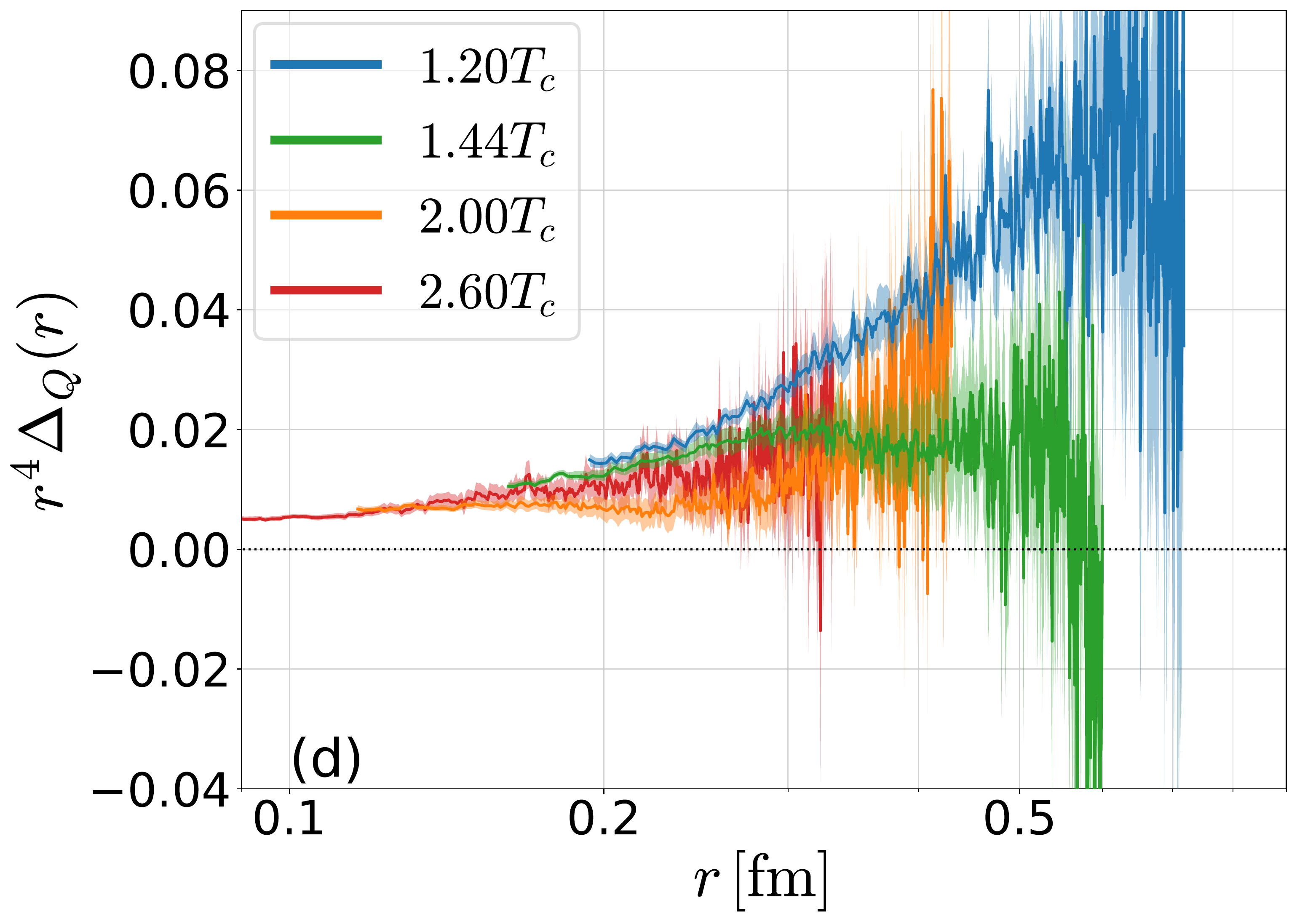}
 \caption{
 EMT distribution 
 $r^4( -\langle T_{44} (r)\rangle_{Q},
 -\langle T_{rr} (r)\rangle_{Q},
 \langle T_{\theta \theta} (r)\rangle_{Q} )$
 and $r^4\Delta_{Q}(r)$
 as functions of $r$~[fm]
 for four temperatures~\cite{Yanagihara:2020tvs}.
 }
 \label{fig:each_temp_r4}
\end{figure}

Next, in Fig.~\ref{fig:each_temp_r4} we show
the distribution of EMT in a different normalization.
Now we show $r^4 \langle T_{\gamma\gamma} (r)\rangle_{Q}$
as a function of $r$~[fm] in physical units.
The distribution of the trace of EMT 
\begin{align}
 \Delta_Q(r) &\equiv -\langle T_{\mu\mu}(r) \rangle_Q 
 = -\langle T_{44}(r)
 + T_{rr}(r) 
 + 2T_{\theta \theta}(r)\rangle_Q.
\end{align}
is also shown in the figure.
From the figure one finds that
the EMT distributions have small $T$ dependence at  
short distances, $r \lesssim 0.2$~fm, while 
sizable $T$ dependence emerges for large distances.
This result is reasonable since the $T$ dependence of 
$r^4 \langle T_{\gamma \gamma'} (r)\rangle_{Q}$ would
be suppressed for $r\lesssim(2\pi T)^{-1}$.

In the leading order perturbation theory in this regime,
we have the following ratio
\begin{align}
 \left|\frac{\Delta_Q(r) }{ \langle T_{44,rr,\theta\theta}(r) \rangle_Q}\right|
 = \frac{11}{2\pi} \alpha_s + \mathcal{O}(g^3),
 \label{eq:alpha}
\end{align}
which is independent of $r$ and $T$ and 
is given only by a function of $\alpha_s$.
The value of $\alpha_s$ estimated from Eq.~(\ref{eq:alpha}) is
$\alpha_s = 0.22-0.32$ depending on the channel at $r=0.1$~fm~\cite{Yanagihara:2020tvs}.

At the long-distance region in Fig.~\ref{fig:each_temp_r4},
owing to the large errors in this region,
it is difficult to extract the thermal screening of
the form $\exp (-2m_{\rm _D} r)$ 
with $m_{\rm _D}$ being the Debye screening mass.
Nevertheless, Fig.~\ref{fig:each_temp_r4} indicates that
the EMT distribution decreases faster than $1/r^4$ in this region,
and the tendency is stronger at  high temperatures. 
To draw a definite conclusion, however, higher statistical data 
are necessary. 

\section{Summary and Concluding remarks}
\label{sec:summary}

In this proceedings
we have studied the EMT distribution around a static quark
in the SU(3) YM theory at finite temperature above $T_c$~\cite{Yanagihara:2020tvs}.
We have used the SF$t$X method based on the gradient flow
to define the EMT operator on the lattice.
We found a substantial difference between 
the EMT distribution in the temporal direction and that of the 
spatial directions, especially near $T_c$.
This separation is a unique feature of the YM theory
that is not observed in the Maxwell theory.

There are interesting future  problems.
First, the extension to full QCD is an important next step.
Although the present analysis in the SU(3) YM theory
was restricted to $T>T_c$,
because the $Z_3$ symmetry is explicitly broken by dynamical fermions,
the present method can be applied directly 
to a single static quark $Q$, as well as static $QQ$ and $Q\bar{Q}$ systems,
both at low and high temperatures.
In particular, the single quark system in QCD at zero temperature
corresponds to a heavy-light meson.
The analysis in this system thus provides us with
the gravitational form factor of the heavy-light meson
at zero and non-zero temperatures.

The numerical simulation of this study was carried out on OCTOPUS
at the Cybermedia Center, Osaka University and Reedbush-U
at Information Technology Center, The University of Tokyo.
This work was supported by JSPS Grant-in-Aid for Scientific Researches, 
17K05442, 18H03712, 18H05236, 18K03646, 19H05598, 20H01903.

\end{document}